\documentclass[showpacs,preprintnumbers,twocolumn,aps,prl,reprint,superscriptaddress,nobibnotes,nofootinbib]{revtex4-1}

\usepackage{amsmath,amssymb,hyperref,times,graphicx,bm,subfigure,bbold,color}
\usepackage{bibunits}
\usepackage{wasysym}

\def\<{\langle}
\def\>{\rangle}

\newcommand{\Xdag}[2]{\hat{{#1}}_{#2}^\dag}
\newcommand{\Xd}[2]{\hat{{#1}}_{#2}^{\phantom\dagger}}

\newcommand{\cdag}[1]{\Xdag{c}{#1}}
\newcommand{\cd}[1]{\Xd{c}{#1}}
\newcommand{\ctdag}[1]{\Xdag{\tilde{c}}{#1}}
\newcommand{\ctd}[1]{\Xd{\tilde{c}}{#1}}
\newcommand{\fdag}[1]{\Xdag{f}{#1}}
\newcommand{\fd}[1]{\Xd{f}{#1}}
\newcommand{\ftdag}[1]{\Xdag{\tilde{f}}{#1}}
\newcommand{\ftd}[1]{\Xd{\tilde{f}}{#1}}

\newcommand{\eq}[1]{Eq.~(\ref{#1})}  
\newcommand{\fig}[1]{Fig.~\ref{#1}}

\newcommand{\JSTAT}{J.~Stat.~Mech.}

\newcommand{\PRL}{Phys.~Rev.~Lett.}

\begin{document}

\defaultbibliographystyle{apsrev4-1_custom}

\title{Kondo Breakdown via Fractionalization in a Frustrated Kondo Lattice Model}

\author{\firstname{Johannes S.} \surname{Hofmann}}
\email{johannes.hofmann@physik.uni-wuerzburg.de}
\affiliation{\mbox{Institut f\"ur Theoretische Physik und Astrophysik, Universit\"at W\"urzburg, Am Hubland, D-97074 W\"urzburg, Germany}}
\author{\firstname{Fakher F.} \surname{Assaad}}
\email{assaad@physik.uni-wuerzburg.de}
\affiliation{\mbox{Institut f\"ur Theoretische Physik und Astrophysik, Universit\"at W\"urzburg, Am Hubland, D-97074 W\"urzburg, Germany}}
\author{\firstname{Tarun} \surname{Grover}}
\email{tagrover@ucsd.edu}
\affiliation{Department of Physics, University of California at San Diego, La Jolla, CA 92093, USA}
\begin{abstract}

We consider Dirac electrons on the honeycomb lattice Kondo coupled to spin-1/2 degrees of freedom on the  kagome lattice.    The interactions  between the spins  are chosen along the lines of the  Balents-Fisher-Girvin model that is known to host a $\mathbb{Z}_2$ spin liquid and a ferromagnetic phase. The model is  amenable to   sign free auxiliary field quantum Monte Carlo simulations.      While in  the ferromagnetic phase the  Dirac electrons acquire a  gap,  they remain massless in the  $\mathbb{Z}_2$ spin liquid phase due to the breakdown of Kondo screening.  Since our model has  an odd number of spins per unit cell,  this phase is a non-Fermi liquid that violates the conventional Luttinger theorem which relates the Fermi surface volume to the particle density in a Fermi liquid. This non-Fermi liquid is a specific realization of the so called fractionalized Fermi liquid  proposed in the context of heavy fermions.  We probe the Kondo breakdown in this non-Fermi liquid phase via conventional observables such as the spectral function,  and also by studying the mutual information between the electrons and the spins.
\end{abstract}

\maketitle

\emph{Introduction:} 
Electron-electron interactions can localize charge carriers and generate insulating states with  local moments \cite{Imada_rev}. What happens when these local moments (f-spins) are  Kondo  coupled  with magnitude $J_K$ to  extended Bloch  conduction (c-) electrons?  For a single local moment, the answer is known:  the Kondo coupling is  relevant and  the f-electron is   screened by the conduction electrons \cite{Anderson70, Hewson}. For a lattice of f-electrons i.e. Kondo lattice systems, the problem is much harder, and the answer is not known in general. However, in the absence of any magnetic ordering, Lieb-Shultz-Mattis-Hastings-Oshikawa theorem \cite{Oshikawa00a, Hasting04, lieb1961} puts strong constraints on the possible outcomes. Specifically, in addition to a heavy Fermi liquid phase where the Fermi surface is `large' since it includes the local moments, there exists a distinct possibility where f-spins decouple from the conduction electrons at low-energies and enter a spin-liquid phase \cite{Senthil03,Senthil04}. In such a `fractionalized Fermi liquid' phase (henceforth denoted as `FL* phase' following Refs.\cite{Senthil03,Senthil04}), the conduction electron Fermi surface is `small' in that it does not include local moments, and therefore the conventional Luttinger theorem \cite{luttinger1960} is violated.

From an experimental standpoint,  a possible breakdown of Kondo screening is relevant to some of the most challenging issues   in  heavy fermion materials \cite{coleman2001, Si01, Senthil03}. There are at least two conceptually different scenarios where a breakdown of Kondo  screening might play a role: in materials such as YbRh$_2$Si$_2$  \cite{Paschen04} and CeCu${}_{6-x}$Au${}_x$ \cite{Klein08}, one observes signatures that indicate that Kondo screening might abruptly change across the transition from a heavy Fermi liquid phase to a magnetically ordered phase. For example, in YbRh$_2$Si$_2$, one observes a jump in the Hall coefficient across the phase transition while in CeCu${}_{6-x}$Au${}_x$, one finds that the single ion Kondo energy scale $T_K$ exhibits an abrupt change close the quantum critical point. A different scenario, which is perhaps more closely related to this paper is the transition from a heavy Fermi liquid to a \textit{non-magnetic} phase across which Kondo screening breaks down. Signatures of such a phase were seen in Co and Ir doped  YbRh$_2$Si$_2$ \cite{Friedmann09}.  Following Refs. \cite{Oshikawa00a,Senthil03,Senthil04}  and as discussed above briefly,  in the absence of any other symmetry breaking (e.g. lattice translation) such a non-magnetic phase is inconsistent with a Fermi liquid ground state if the Kondo screening is not operative and the unit cell contains an odd number of spin-1/2 spins. The local moments in such a phase are then forced to either have a gapless spectrum or topological order \cite{Hasting04}.  We also note that as discussed in Ref.~\cite{Vojta10}, the Kondo breakdown is also closely related to the concept of `orbital selective Mott transition'. In addition, there are several other heavy fermionic materials such as CePdAl \cite{doenni96,goto02,oyamada08,akito16}, $\kappa$-(ET)${}_4$Hg${}_{2.89}$Br${}_8$ \cite{oike2016}, YbAgGe \cite{kim_frustkondo08}, YbAl${}_3$C${}_3$ \cite{sengupta_frustkondo10} and Yb${}_2$Pt${}_2$Pb \cite{kato_frustkondo08} whose phenomenology seems to be poorly understood, and where microscopic considerations suggest that the geometric frustration between local moments plays an important role.

\begin{figure}
	\centering
	\includegraphics[width=1.\linewidth]{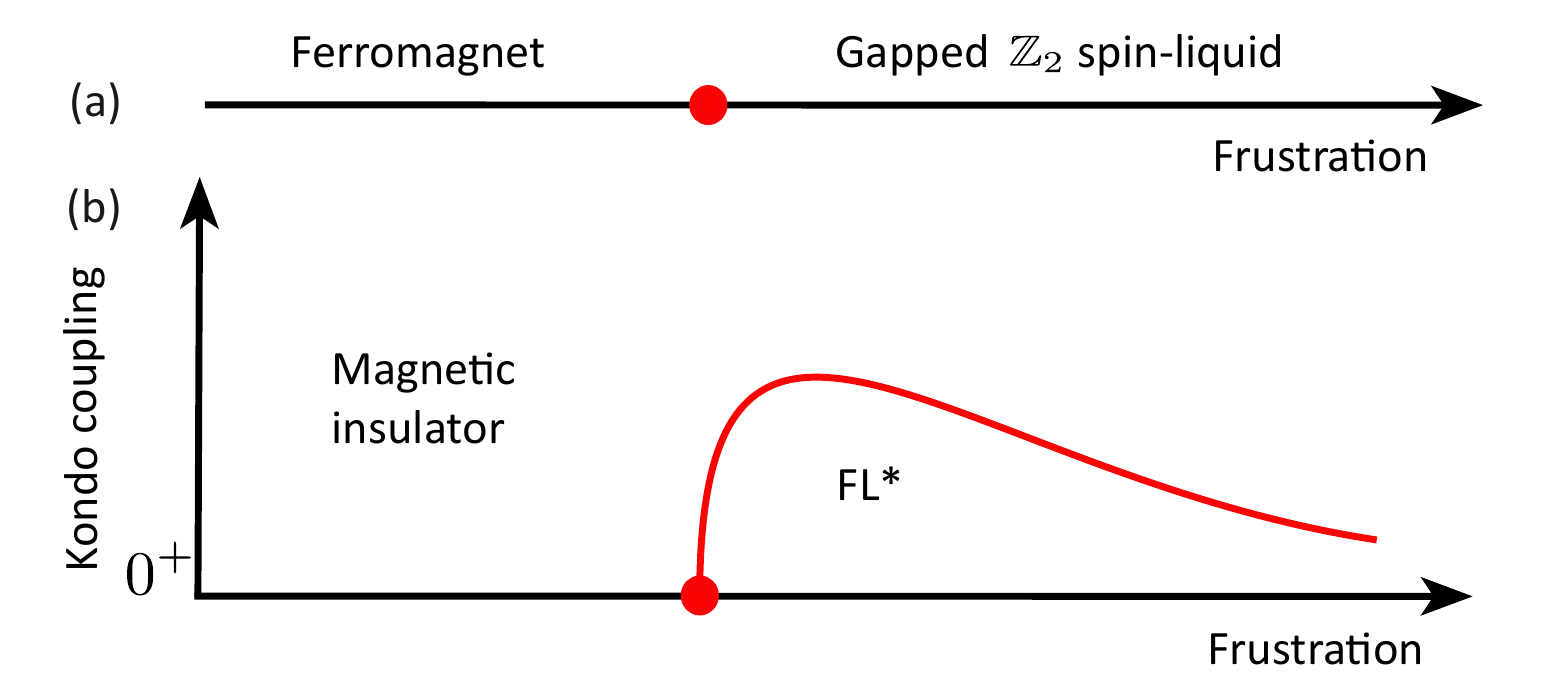} 
	\caption{ (color online) (a) Schematic phase diagram of the BFG model in the absence of Kondo coupling. (b) Schematic phase diagram of the BFG model in the presence of Kondo coupling.
	}
	\label{fig:schematic}
\end{figure}

In this paper we will introduce a generalized  Kondo lattice model which hosts the aforementioned Kondo breakdown transition 
between a conventional phase with electron like quasiparticles, and an FL* phase with $\mathbb{Z}_2$ topological order.  From a technical standpoint, the most salient feature of our model is that it does not suffer from fermion sign problem even in the presence of the Kondo coupling \cite{SatoT17_1}. Our model is realized by Kondo coupling a variant of the Balents-Fisher-Girvin  (BFG) model  \cite{Balents02,Isakov06, Isakov07}, first introduced in 
Ref.~\cite{IHM-11}, to conduction electrons. The BFG model supports a transition from a ferromagnetic phase to a  gapped $\mathbb{Z}_2$ spin-liquid (Fig. \ref{fig:schematic}(a)). When this model is weakly coupled to conduction electrons, the spin-liquid gives way to an FL* phase where the conduction electrons form a Dirac semi-metal, while the local moments continue to form a $\mathbb{Z}_2$ spin-liquid (Fig.\ref{fig:schematic}(b)).    Since our unit cell contains two c-electrons and three f-spins,   this  result stands at odds with the Luttinger sum rule. As the Kondo coupling is increased beyond a threshold, one loses the topological order of local moments, and enters a conventional phase with electron like quasiparticles. We will characterize the Kondo breakdown by studying the spectral function of the conduction electrons, and also via the mutual information between the conduction electrons and local moments.

\begin{figure}
	\centering
	\includegraphics[width=1.\linewidth]{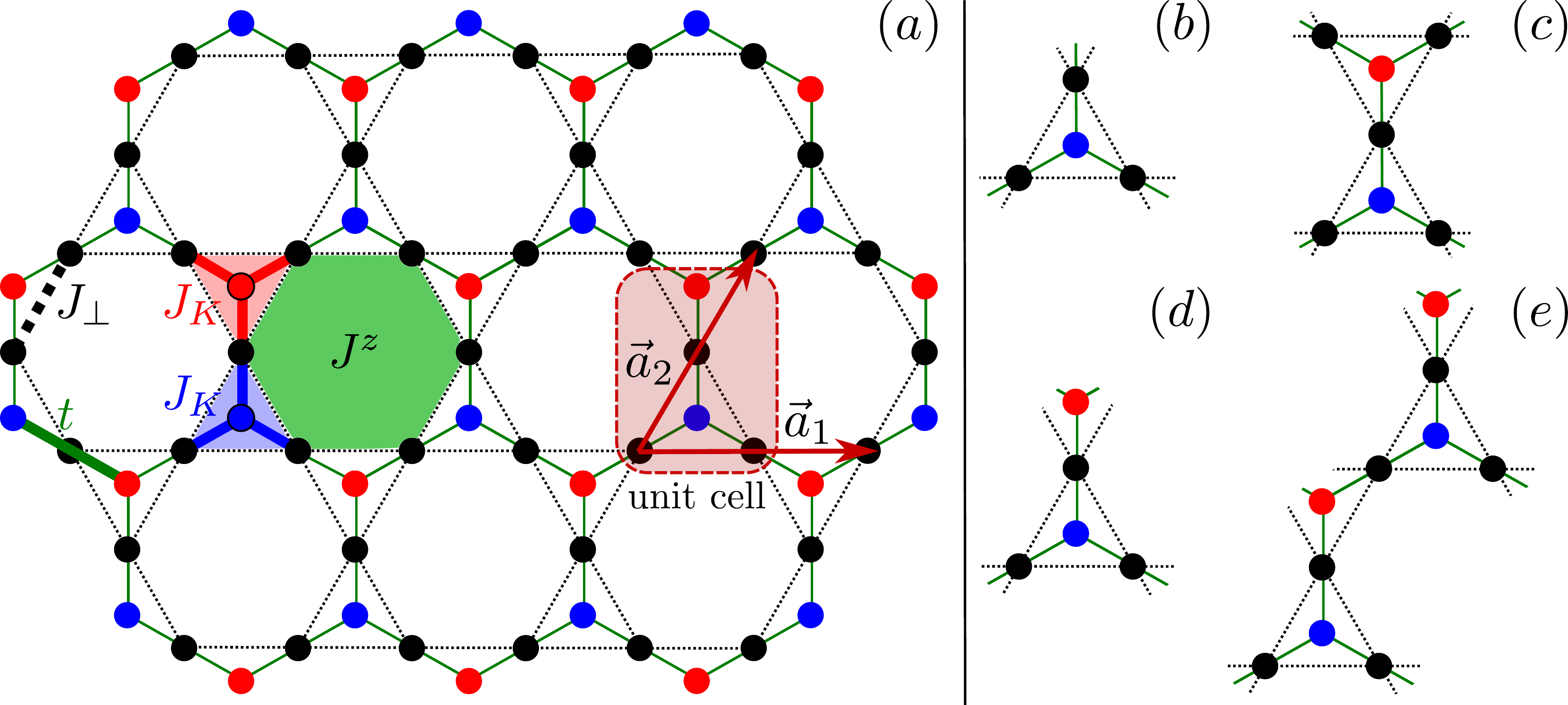} 
	\caption{ (color online) \emph{Left:} The model - The conduction (c-) electrons   hop, with matrix element $t$,  between  nearest neighbor sites of the honeycomb lattice  denoted by the red and blue circles. The kagome  lattice (black) supports impurity spins described by the Balents-Fisher-Girvin model with nearest neighbor spin-flip $J^\perp$ and interactions on hexagons of strength $J^z$ (green). The two  systems are Kondo-coupled with strength $J_K$ for each bond in the elemental triangles (thick red and blue bonds). For details see \eq{KondoH}.
	\emph{Right:} Various patches $\Gamma$ used to extract the Renyi mutual information. Subsets (b) and (c) belong to the triangle sequence, (d) and (e) are built out of unit cells.
	}
	\label{fig:Lattice}
\end{figure}

\emph{Model and limiting cases:} 
We investigate the following generalized Kondo lattice model (KLM) described by $\hat{H} =  \hat{H}_c +  \hat{H}_S + \hat{H}_K$ with:
\begin{eqnarray}
\label{KondoH} 
\hat{H} _c & =  & 
-t \sum_{\< \pmb{x},\pmb{y}\> ,\sigma}   \ \cdag{\pmb{x},\sigma}  \cd{\pmb{y},\sigma} + h.c.\\
\hat{H}_{\mathrm{S}}&=&-J^\perp\sum_{\left\langle \pmb{i},\pmb{j} \right\rangle}\left( \hat{S}^{f,+}_{\pmb{i}}   \hat{S}^{f,-}_{\pmb{j}} + h.c.  \right) +J^z \sum_{\hexagon} \left(\hat{S}^{f,z}_{\hexagon} \right) ^ 2
\nonumber \\
\hat{H}_K &= & J_K \sum_{\< \pmb{x}, \pmb{i}\>} \left[ \hat{S}^{c,z}_{\pmb{x}}   \hat{S}^{f,z}_{\pmb{i}} - (-1)^{\pmb{x}} \left( \hat{S}^{c,+}_{\pmb{x}}   \hat{S}^{f,-}_{\pmb{i}} + h.c. \right) \right].
\nonumber   
\end{eqnarray}
Here, $ \cdag{\pmb{x},\sigma} $ creates a conduction electron in a Wannier state  centered at $\pmb{x}$ with a z-component of spin $\sigma$,   $ \pmb{S}^{c}_{\pmb{x} }  = \frac{1}{2} \sum_{s,s'}   \cdag{\pmb{x},s}  \bm{\sigma}_{s,s'} \cd{\pmb{x},s'}$ is the spin operator  and $\< \pmb{x},\pmb{y}\>$ are the nearest neighbors of a honeycomb lattice.   $ \pmb{S}^{f}_{\pmb{i} }$  is a spin-1/2 degree of freedom  located on the kagome lattice   corresponding to the  median of the honeycomb lattice (see Fig.\ref{fig:Lattice}). The Hamiltonian $\hat{H}_S$  is a variant of the BFG model (Ref.~\cite{Balents02,IHM-11}) with nearest neighbor, $\< \pmb{i},\pmb{j}\>$, spin flip  amplitude $J^\perp$ and   interaction, $J^z$  that  minimizes the total z-component of spin on a hexagon: $\hat{S}^{f,z}_{\hexagon}=\sum_{\pmb{i}\in\hexagon}\hat{S}^{f,z}_{\pmb{i}}$.
The conduction electrons and the local moments are Kondo coupled,  according to  $\hat{H}_K$,  along nearest neighbor bonds $\< \pmb{x},\pmb{i}\>$ between the kagome and Honeycomb lattices (Fig. \ref{fig:Lattice}). The factor  $(-1)^{\pmb{x}}$ that takes the value $1$ ($-1$) on the A (B) sublattice of the Honeycomb lattice is necessary to avoid the negative sign problem.   In particular it cannot be gauged away since the  kagome lattice is not bipartite. Referring back to Fig.\ref{fig:schematic}, $J^z$ plays the role of frustration, and $J_K$ is the Kondo coupling.

Let us consider various limiting cases of the Hamiltonian $\hat{H}$. When $J^\perp \gg J^z, J_K$, the local moments order in an an $XY$-ferromagnetic ground state.  Taking into account the $ (-1)^{\pmb{x}} $ factor in the Kondo coupling, we see  that this terms induces an anti-ferromagnetic in-plane mass term for the conduction electrons.  Hence, in this limit one obtains a magnetically ordered insulating phase.

Next, consider $J_K \gg J^\perp \gtrsim J^z, t$. First, let us set all couplings except $J_K$ to zero. Performing the unitary transformation $\cd{\pmb{x},\downarrow}\rightarrow -(-1)^{\pmb{x}} \cd{x,\downarrow}$ maps the Kondo interaction to an anti-ferromagnetic Heisenberg coupling between the conduction electrons and the local moments. This interaction is not frustrated, and the ground state is AFM ordered with opposite polarizations on the kagome sites and the Honeycomb lattice. Undoing  the above transformation, the in-plane magnetization of the conduction electrons will be parallel for one honeycomb sublattice and anti-parallel for the other, relative to the local moments. Next, turning on a small $J^\perp, J^z$ with $J^\perp \gtrsim J^z$, the local moments will preferably order in the XY plane. Comparing to the limit  $J^\perp \gg J^z, J_K$, one finds that the in-plane symmetry breaking pattern is identical and in the absence of any out-of-plane component, this phase is expected to be adiabatically connected to the aforementioned magnetically ordered insulating phase in the $J^\perp \gg J^z, J_K$ limit. Note that an out-of-plane component will spontaneously break the symmetry $\hat{S}^{f,z}_{\pmb{i}} \rightarrow - \hat{S}^{f,z}_{\pmb{i}}, \hat{S}^{f,x}_{\pmb{i}} \rightarrow  \hat{S}^{f,x}_{\pmb{i}}, \hat{S}^{f,y}_{\pmb{i}} \rightarrow  \hat{S}^{f,y}_{\pmb{i}}$ (see the supplemental material for a detailed discussion of the symmetries). Due to symmetry breaking and associated stiffness,  this phase is stable also to switching on a small hopping $t$.

Most interesting is the limit $J^z \gg J^\perp \gg J_K$. When only $J^z$ and $t$ are non-zero, the conduction electrons form a Dirac semimetal while the local moments can be described as  a classical system with a ground state degeneracy that scales exponentially with the system size \cite{Balents02}. Allowing a small $J^\perp/J^z \ll 1$ lifts this macroscopic degeneracy and leads to a $\mathbb{Z}_2$ topologically ordered spin liquid of the local moments \cite{Balents02}. Remarkably, as discussed in  Refs. \cite{Senthil03,Senthil04}, introducing a small  Kondo coupling $J_K$ leaves the state unchanged because perturbatively the Kondo coupling is irrelevant at the renormalization group fixed point where conduction electrons form a Dirac semimetal while the local moments are in a gapped  $\mathbb{Z}_2$ topologically ordered state. Therefore, at low energies, the local moments decouple from the conduction electrons and one obtains a non-Fermi liquid FL* phase with a `small' Fermi surface which was introduced in   Refs.\cite{Senthil03,Senthil04}. Physically, in this phase the local moments are highly entangled with each other such that the formation of Kondo singlets or the tendency to magnetically order is suppressed.

The phases discussed above, especially the FL* phase, should be contrasted with the conventional heavy Fermi liquid that satisfies the  Luttinger sum rule. Since our model has two electrons and three spins per unit  cell, the most prominent feature is that this state has a `large' Fermi surface which encloses half of the BZ whereas the Fermi volume of the aforementioned fractionalized FL* phase vanishes.  The nature of the  Fermi liquid state  strongly depends on symmetries.  If particle hole-symmetry (PHS) is imposed in the paramagnetic phase, then one would expect a  flat-band pinned at the Fermi level, a generically unstable state \cite{Derzhko15,Honerkamp2000,Potter2014,Hofmann16,Lieb89,Bercx17,Feldner11,graphene_edge_magnetism,Tang14}. A hybridization between $c$- and $f$-electrons necessarily breaks either PHS - with uniform hybridization - or TRS - when the $(-1)^{\pmb{x}}$  phase in the  Kondo coupling is  carried over to the  hybridization. The latter requires fine-tuning to remain paramagnetic whereas the former can generate a non-magnetic heavy Fermi liquid. In the range of parameters considered in this paper, we do not find such a phase.
A more detailed discussion can be found in the supplemental material.

\emph{Method and observables:}  
We simulate the Hamiltonian in Eq.~(\ref{KondoH}) using the auxiliary field quantum Monte Carlo (QMC) method \cite{Blankenbecler81,White89,Assaad08_rev}. We follow the strategy outlined in Ref.~\cite{SatoT17_1} where it was shown that Hamiltonians of the form $\hat{H}$ do not suffer from fermion sign problem when $J^{\perp}  \geq 0$ and the conduction bands are particle-hole symmetric.  In this approach local moments are fermionized, $ \pmb{S}^{f}_{\pmb{i} }  = \frac{1}{2} \sum_{s,s'}   \fdag{\pmb{i},s}  \bm{\sigma}_{s,s'} \fd{\pmb{i},s'}$,  
with the constraint $\sum_{s}\fdag{\pmb{i},s} \fd{\pmb{i},s} = 1$.  As in simulations of the generic  Kondo lattice model \cite{Assaad99a,Capponi00}   this constraint can be imposed very efficiently since it corresponds to a local conservation law.  The details of our implementation are summarized in the  supplemental  material and we have used the ALF package \cite{ALF} to carry out the simulations.  
Despite the absence of sign problem, the simulations of this model  are challenging.   Fermionization   leads to a large number of  auxiliary fields (33 per unit cell),  and the {\it condition number}  on scales   corresponding to the ratio of {\it band width}  to the    smallest relevant scale (e.g.  vison gap in the $\mathbb{Z}_2$ spin liquid phase) is  large.  As a consequence, we have used an imaginary time step $\Delta \tau t = 0.01$.    The biggest challenge turns out to be large autocorrelation times. We tried to improve this issue by using global moves that mimic vison excitations,  as well as by  implementing  parallel tempering schemes.  Nevertheless, these long autocorrelation times remain the limiting factor  to  access system sizes bigger than those presented  here, in particular $3 \times 3$ and  $ 6 \times 3 $ unit cells.  For both lattices sizes, and the considered periodic boundary conditions,  Dirac points are present. However, only the $ 6 \times 3 $  allows to satisfy $\hat{S}^{f,z}_{\hexagon}=0$ for all hexagons. 

We compute  spin-spin correlations $S_{AFM}= 1/L \sum_{\pmb{I}\pmb{J}}\< \hat{S}^x_{\pmb{I}} \hat{S}^x_{\pmb{J}} + \hat{S}^y_{\pmb{I}} \hat{S}^y_{\pmb{J}} \>$  where the net  spin per unit cell $\pmb{I}$,   $\hat{\pmb{S}}_{\pmb{I}}=\sum_{\pmb{i}\in \pmb{I}}\hat{\pmb{S}}^{f}_{\pmb{i}} + \sum_{\pmb{x}\in \pmb{I}}(-1)^{\pmb{x}}\hat{\pmb{S}}^{c}_{\pmb{x}}$,  captures the  aforementioned ferromagnetic-antiferromagnetic order of the f-spins and conduction electrons.     The  spectral function of the conduction electrons $A_{c}(\pmb{k},\omega)
= - \frac{1}{\pi} \text{ Im }G^{\text{ret}}_{c}( \pmb{k},\omega )$   can be  extracted from the imaginary time resolved Greens function $G_c(\pmb{k},\tau)=\sum_{\alpha,\sigma}\<\cdag{\pmb{k},\alpha,\sigma}(\tau)\cd{\pmb{k}, \alpha, \sigma}(0)\>$ using the MaxEnt method \cite{Sandvik98,Beach04a}. Here $\alpha$  is the orbital index.
The auxiliary field QMC method also allows to study the entanglement properties of fermionic models \cite{Peschel-04,Grover-13,ALPT-13,DP-15,DP-16,PTA-18}. In particular, as shown in Refs.~\cite{Grover-13,ALPT-13},  the second Renyi entropy $S_2$ can be computed from the knowledge of Greens-functions $G_A$, restricted to subsystem $A$ for two independent Monte Carlo samples.   An alternative  approach exploits the replica trick, e.g. for fermionic \cite{BT-14,WT-14,Assaad-15,BT-16}, bosonic \cite{IHM-11}, and spin systems \cite{HGKM-10,HR-12}. For a  given  subsystem of conduction electrons   $\Gamma_{c}$ and  of  spins $\Gamma_{f}$,  the Renyi mutual   information  between  $\Gamma_{c}$ and $\Gamma_{f}$ is
$I_2(\Gamma_c, \Gamma_f)     \equiv  S_2(\Gamma_{c}  \cup \Gamma_{f} ) - S_2(\Gamma_{c}) - S_2(\Gamma_{f} )$. We use the two sequences for $\Gamma$ as shown  in  \fig{fig:Lattice}(b), (c) and,  \fig{fig:Lattice}(d),(e).  In the  calculation of the Renyi mutual information  we restore the $C_3$ lattice symmetry by  averaging over rotationally  equivalent $\Gamma$s.

\emph{Results:} 
From here on, we fix $J^\perp=t$ and use $t=1$ as the unit of energy.   The BFG model shows a transition from the ferromagnetic state to the 
$\mathbb{Z}_2$ spin liquid at $J^{z}_c  \simeq  7.07 $ \cite{IHM-11}.   Alongside with spin excitations, the $\mathbb{Z}_2$  spin liquid hosts vison excitations.    Recent simulations of  the   dynamics of the BFG model \cite{Becker18}  estimate the spin and  vison gaps at $J^z  =  8.\overline{3}$ to $\Delta_s \simeq 7.12 $ and  $\Delta_v \simeq 0.2 $.    We expect that the  vison gap remains non-zero at the transition and that the spin gap scales as  $   \left( J^z - J^z_c \right)^{\nu z} $ with dynamical critical exponent  $z=1$ and $\nu \simeq 0.67 $, which correspond to the exponents of the 3D XY* model \cite{Chubukov94_2, Isakov06,Grover10a}.

\begin{figure}
	\centering
	\includegraphics[width=1.\linewidth]{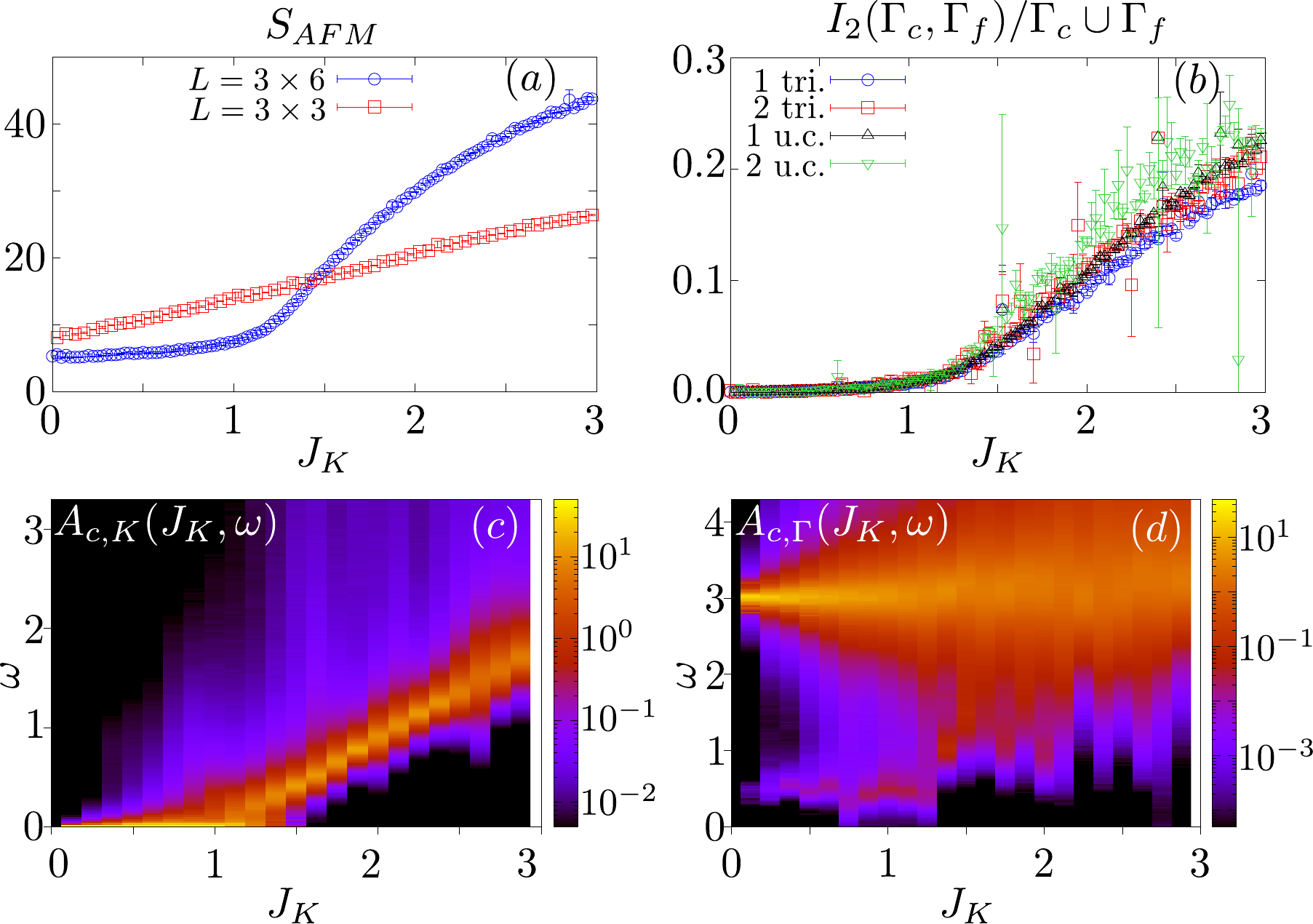}
	\caption{ (color online)
		We consider  lattices $L=3\times 3$ and $L=3\times 6$  unit cells  at an inverse temperature $\beta =12$ and at $ J^z  = 7.5 $ 
		 (a)  Spin-spin  correlations  $S_{AFM}$ (See text),  (b) Renyi mutual informations $I_2(\Gamma_c, \Gamma_f)$ per site of the patch  $\Gamma_{c}  \cup \Gamma_{f}$ for $L=3\times 6$.   Here we consider the patches listed in \fig{fig:Lattice}(b)-(e). 
		 (c)   Conduction electron spectral function  at the Dirac point $\pmb{K}$   for the $3 \times 6 $ lattice. (d) Same as (c), but at the $\Gamma$-point.  
		 The imaginary time data from which panels (c) and (d) stem are presented in the supplemental material.  }
	\label{fig:Z2_Jscan}
\end{figure}

\fig{fig:Z2_Jscan}   shows  a scan at $J^z  = 7.5$ as a function of $J_K$.  We  have set the temperature to $\beta  = 12$.  From the above 
discussion,   this choice of temperature places us well below the spin gap and  allows us to resolve the vison gap.  
As apparent in  \fig{fig:Z2_Jscan}(c), the single particle spectral function at the Dirac point   remains gapless. As a function of  $J_K$ it looses spectral weight and a full gap opens    sightly  before $J_K = 1.5$.     At this energy scale the spin-spin correlations  $S_{AFM}$  show a marked upturn (see  \fig{fig:Z2_Jscan}(a)).  In the presence of long ranged magnetic order $S_{AFM}$  scales as the volume of the system. Comparison between the $3\times 3 $   and $ 3 \times 6$ lattices shows that  $S_{AFM}$  grows as a function of system size beyond $J_K=1.5$.   

\begin{figure}
	\centering
	\includegraphics[width=1.\linewidth]{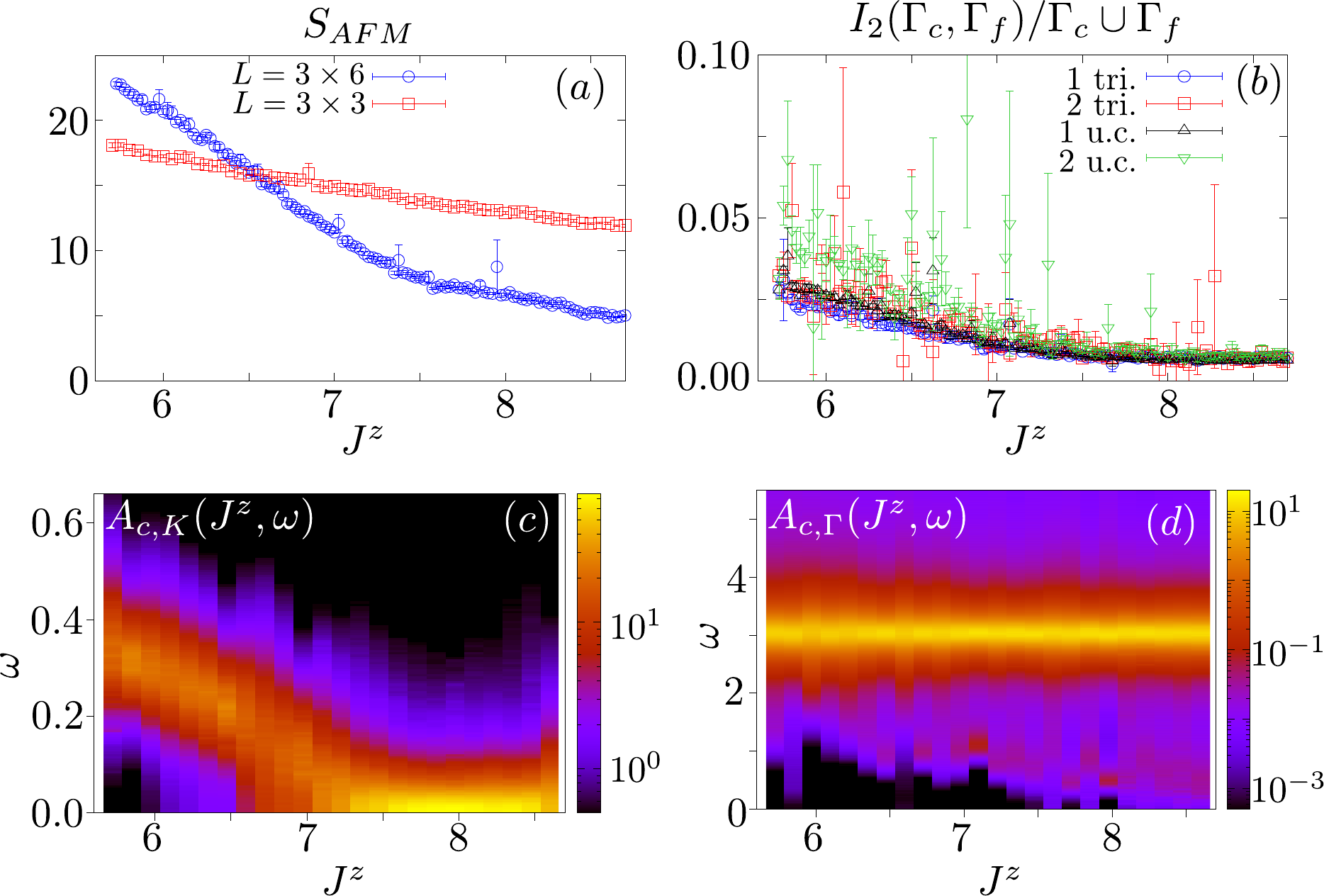}
	\caption{ (color online)
		We consider  lattices $L=3\times 3$ and $L=3\times 6$  unit cells  at an inverse temperature $\beta =12$ and at $ J_K  = 1 $ 
		 (a)  Spin-spin  correlations  $S_{AFM}$ (See text),  (b) Renyi mutual informations $I_2(\Gamma_c, \Gamma_f)$ per site of the patch  $\Gamma_{c}  \cup \Gamma_{f}$ for $L=3\times 6$.   Here we consider the patches listed in \fig{fig:Lattice}(b)-(e). 
		 (c)   Conduction electron spectral function  at the Dirac point $\pmb{K}$   for the $3 \times 6 $ lattice. (d) same as (c), but at the $\Gamma$-point.   The imaginary time data from which panels (c) and (d) stem are presented in the supplemental material.  }
	\label{fig:Z2_Vscan}
\end{figure}

Small values of $J_K$  are associated with small energy  scales which  may be difficult to resolve on our finite sized systems  at finite temperatures.  
To confirm  above result, we  present  a scan at fixed $J_K = 1$   and  vary $J^z$ in  \fig{fig:Z2_Vscan}.  
Upon analysis of Figs.~\ref{fig:Z2_Vscan}(a) and \ref{fig:Z2_Vscan}(c) one  concludes that the  magnetic order and the single particle  gap track each other.  In particular the single particle gap closes in the  $\mathbb{Z}_2$ spin liquid phase. 

Signatures of the $\mathbb{Z}_2$ spin liquid phase can be picked up in the spectrum of the conduction electrons.  In \fig{fig:Z2_Jscan}(d) and \fig{fig:Z2_Vscan}(d)  we plot the single particle spectral function at the $\Gamma$ point.  One notices that in the FL* phase, spectral weight at low energies is apparent.  We associate this  feature with the vison excitations of the  $\mathbb{Z}_2$ spin liquid.

It is interesting to consider other measures for Kondo screening. The Renyi mutual information $I_2$ between the c-electrons and the f-spins introduced above provides one such measure. It is   important to note that this quantity is both IR and UV sensitive since we are considering mutual information between two Hilbert spaces that overlap in  real space. Despite the decoupling of conduction electrons and local moments at low energies in the FL* phase, one therefore doesn't except that the mutual information will be exactly zero in this phase. It vanishes only at the RG fixed point corresponding to $J_K = 0$, where these two Hilbert spaces completely decouple.  In the opposite limit when the c-electrons and f-spins are maximally entangled, the Renyi mutual information will attain its maximum possible value of $4 \log(2)/5$ per site (recall that the unit cell of our model contains three  f-spins and  two c-electrons ). In the magnetically ordered phase, one expects that the Renyi mutual information will not be close to this maximum due to the entanglement between the local moments themselves. From  \fig{fig:Z2_Jscan} (b) and \fig{fig:Z2_Vscan} (b)  we see that the QMC data is consistent  with this expectation.  The most notable  feature is that the Renyi mutual information per site is an order of magnitude smaller in the FL* phase compared to the magnetically ordered phase. Furthermore, even on a limited size lattices such as ours, one can already see signatures of the transition from the magnetically ordered phase to the FL* phase as evidenced by the change of slope in the coefficient of the Renyi mutual information at the transition.

\emph{Conclusion and discussion:}    In this paper we introduced a model  amenable  to negative sign free Monte Carlo simulations that can host a fractionalized Fermi liquid (FL*) phase. The most prominent feature of this   phase is a violation of the Luttinger theorem due to the onset of  topological order.   This proof of principle calculation  paves the way to many other investigations.  We have considered a model where the fractionalization  inherent to topological order is `emergent' i.e. the lattice model is written in terms of spins.  A different, and possibly  numerically more tractable approach would be to simulate  directly a theory of spinons coupled to $\mathbb{Z}_2$ gauge  fields following Refs.~\cite{Assaad16,Gazit16,Gazit18}  and where spinons are also Kondo coupled  to conduction electrons. Such an approach might be particularly useful for studying the quantum phase transition between the FL* phase and the magnetically ordered phase. A field theory description of this transition was provided in Ref.\cite{Grover10a} where it was found that the Kondo coupling is irrelevant at the critical point due to the large anomalous exponent of the spins, and therefore one expects that the conduction electrons have a well defined electron-like quasiparticle even at the critical point, while the local moments will inherit the critical exponents of the 3D XY* transition \cite{Chubukov94_2, Isakov06}.

It might be also interesting to explore the possibility of obtaining non-trivial symmetry protected topological phases in frustrated Kondo models along the lines of Ref.~\cite{yang2017} where it was shown that under certain conditions, one can obtain symmetric states without any topological order even when the unit cell contains an odd number of spins but the magnetic unit cell has an integral number of spins.

 Another avenue to explore would be the universal subleading contribution of the Renyi entanglement entropy for a spatial bipartition. In the FL* phase one expects that this contribution is given as  $\gamma = \gamma_{\textrm{topo}} + \gamma_{\textrm{Dirac}}$, where  $\gamma_{\textrm{topo}} = \log(2)$ is the topological entanglement entropy corresponding to the topological order of the local moments, while $\gamma_{\textrm{Dirac}}$ is the shape-dependent universal contribution from the Dirac conduction electrons \cite{Swingle12, Yao10}. Similarly, at the transition, owing to the aforementioned irrelevance of the Kondo coupling, one expects that $\gamma = \gamma_{\textrm{topo}}+ \gamma_{\textrm{Dirac}}  + \gamma_{\textrm{3D XY}} $ where  $\gamma_{\textrm{3D XY}}$ is the universal shape-dependent entanglement contribution at the 3D XY transition \cite{Swingle12}. 
 
 Finally, as mentioned in the introduction, the Kondo breakdown scenario is central to several questions in heavy fermion materials as well as newly discovered frustrated Kondo lattice systems. Our approach opens a window to quantitatively explore these and related questions as well.

\begin{acknowledgments}
	\emph{Acknowledgments:} 
	The authors thank T. Sato, F. Parisen-Toldin for stimulating discussions and S. Sachdev, A. Vishwanath for comments on the draft. JH and FFA are supported by the German Research Foundation (DFG), under DFG-SFB~1170 ``ToCoTronics" (Project C01).  TG is supported by the National Science Foundation under Grant No. DMR-1752417, and  as an Alfred P. Sloan Research Fellow.
	The authors gratefully acknowledge the Gauss Centre for Supercomputing e.V. (www.gauss-centre.eu) for funding this project by providing computing time on the GCS Supercomputer SuperMUC at Leibniz Supercomputing Centre (LRZ, www.lrz.de). We also acknowledge the Bavaria California Technology Center (BaCaTeC) for travel  support.
\end{acknowledgments}

\bibliography{./francesco,./fassaad}

\clearpage

\appendix

\begin{center}
	\textbf{
		\large{Supplemental Material for}}
	\vspace{0.4cm} 
	
	\textbf{
		\large{
			``Kondo Breakdown via Fractionalization in a Frustrated Kondo Lattice Model" } 
	}
\end{center}

\vspace{0.1cm}

\begin{center}
	\textbf{Authors:}  
	Johannes S. Hofmann,
	Fakher F. Assaad \\
	and Tarun Grover
\end{center}

\section*{I.~~~Symmetries and Heavy Fermi Liquids}

\begin{table}
	\begin{tabular}{|c||c|c|c|c|} 
		\hline
		$U$ & $\sigma^x$ & $\sigma^z$ & $\sigma^0$ & $\sigma^0$ \\ 
		\hline
		$\alpha$ & $-$ & $-$ &  $+$ & $-$ \\ 
		\hline
		$\beta_x$ & $-$ & $+$ &  $-$ & $-$ \\ 
		\hline
		$\beta_y$ & $+$ & $+$ & $+$ & $-$ \\ 
		\hline
		$\beta_z$ & $+$ & $-$ & $-$ & $-$ \\ 
		\hline
	\end{tabular} 
	\caption{Table of independent particle-hole symmetries. See the supplemental material text for the notation. \label{Tab:PHsym}}
\end{table}

Our model, Eq.(\ref{KondoH}), has several continuous and discrete symmetries. Among continuous symmetries, the number of conduction electrons is conserved, and so is the projection of the total spin along the z-direction, i.e., $\sum_{\pmb{x}}\hat{S}^{c,z}_{\pmb{x}} + \sum_{\pmb{i}} \hat{S}^{f,z}_{\pmb{i}}$. 

The model also exhibits several unitary and anti-unitary particle-hole symmetries which we list in Table \ref{Tab:PHsym}. They are implemented by a matrix $U$ via $\cdag{\pmb{x},s}\rightarrow (-1)^{\pmb{x}} U_{s,s'} \cd{\pmb{x},s'}$ and $\fdag{\pmb{i},s}\rightarrow U_{s,s'} \fd{\pmb{i},s'}$ together with the sign $\alpha$ distinguishing between unitary and anti-unitary transformations: $\sqrt{-1} \rightarrow \alpha \sqrt{-1}$. We list their action on the spin operators by the signs $\pmb{\beta}=(\beta_x,\beta_y,\beta_z)$ with $\hat{S}^{c,l}_{\pmb{x}}\rightarrow\beta_l\hat{S}^{c,l}_{\pmb{x}}$ as well as $\hat{S}^{f,l}_{\pmb{i}}\rightarrow\beta_l\hat{S}^{c,l}_{\pmb{i}}$.

One can also combine the particle-hole symmetries in Table \ref{Tab:PHsym} to define two different anti-unitary time-reversal symmetries.  The first one, $TR_1$, is defined via $\cdag{\pmb{x},s}\rightarrow i \sigma_{s,s'}^y \cdag{\pmb{x},s'}$ and $\fdag{\pmb{i},s}\rightarrow i \sigma_{s,s'}^y \fdag{\pmb{i},s'}$ along with $\sqrt{-1} \rightarrow - \sqrt{-1}$. This transformation flips all three components of the spin operators $\hat{\pmb{S}}^{c}_{\pmb{x}}\rightarrow-\hat{\pmb{S}}^{c}_{\pmb{x}}$ as well as $\hat{\pmb{S}}^{f}_{\pmb{i}}\rightarrow-\hat{\pmb{S}}^{f}_{\pmb{i}}$. The second one, $TR_2$, replaces $i\sigma^y$ by $\sigma^x$ so that only the $z$-component of the spin operators gets reversed.

At the level of free fermion band-structure, the particle-hole symmetries listed above lead to flat bands. In particular, either of the symmetries $(U,\alpha)=(\sigma^0,-)$ and $(U,\alpha)=(\sigma^z,-)$  guarantee that there is a flat band. This is because these transformations do not mix up and down spin components, which leads to an odd number (=five) of bands for each spin sector. Furthermore, the anti-unitary nature of the symmetry implies that $c(k) \rightarrow c^{\dagger}(k)$. Thus there always exists a flat band at zero energy in each spin sector. Such a flat band will generically be unstable to interactions, e.g., according to Table~\ref{Tab:PHsym}, a magnetically ordered state in the $z$-direction will break both of these particle-hole symmetries.

\begin{figure}
	\includegraphics[width=0.914\linewidth]{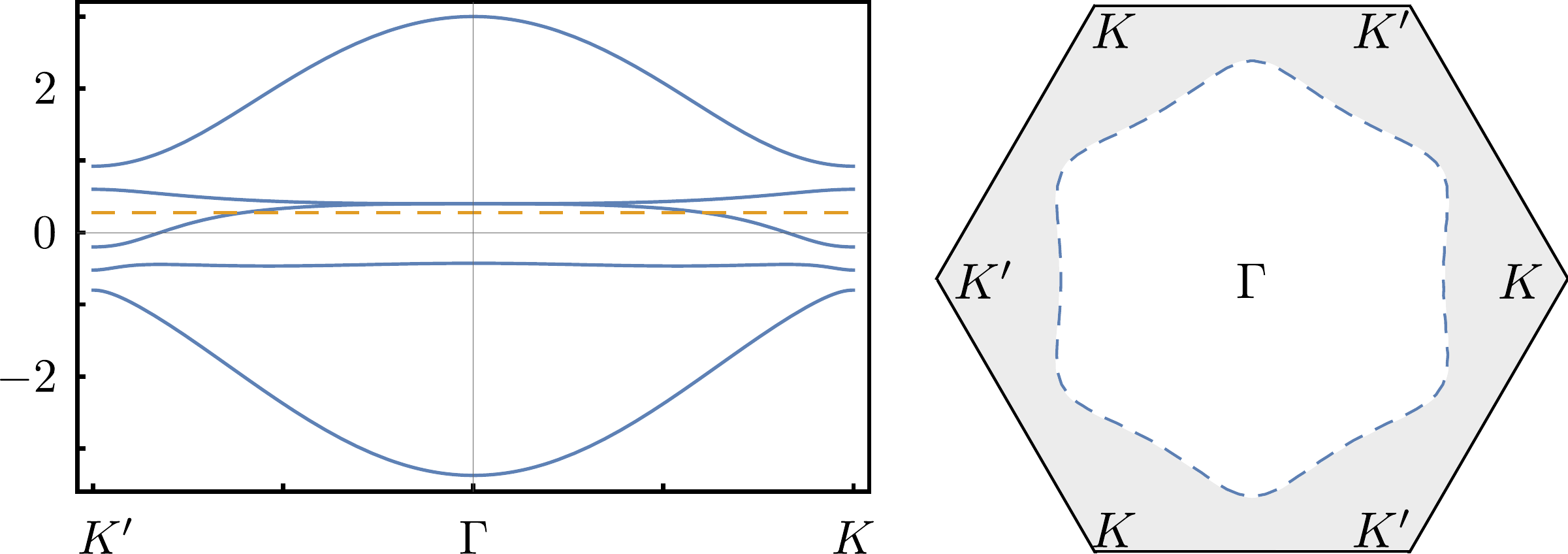}
	\caption{(color online) Cut of the spectrum from $K'$ to $K$ with the Fermi energy marked by the dashed, orange line constraint to the half filled case and Fermi surface (blue, dashed) of a heavy Fermi Liquid state with $\Delta=0.4$ and $t'=0.2$. The shaded area marks the occupied part of the BZ.}
	\label{fig:fullbandstructure}
\end{figure}

Let us next consider heavy fermion phases that result from the hybridization of $c$- and $f$-electrons. As just discussed, to obtain dispersive bands one would need to break at least some symmetries. One option is  a uniform hybridization, $\sum_{\< \pmb{x}, \pmb{i}\>,s} \cdag{\pmb{x},s}\fd{\pmb{i},s} +h.c.\,$, that preserves the time reversal symmetries $TR_1$ and $TR_2$, but breaks all the particle-hole symmetries. As a consequence, one should also allow direct f-electron hopping terms given by $\sum_{\< \pmb{i}, \pmb{j}\>,s} \fdag{\pmb{i},s}\fd{\pmb{j},s} +h.c.\,$. The mean-field Hamiltonian for such a heavy Fermi Liquid is then given as

\begin{equation}
	\hat{H}_{\mathrm{hFL}}=\hat{H}_c+\Delta\sum_{\< \pmb{x}, \pmb{i}\>,s} \cdag{\pmb{x},s}\fd{\pmb{i},s}+t'\sum_{\< \pmb{i}, \pmb{j}\>,s} \fdag{\pmb{i},s}\fd{\pmb{j},s} +h.c.\,.
\end{equation}
The resulting band-structure is depicted in \fig{fig:fullbandstructure} for $\Delta=0.4$ and $t'=0.2$, where the left hand side shows a cut from $K'$ to $\Gamma$ to $K$. We clearly recognize a dispersive band in the middle of the spectrum replacing the aforementioned flat band at zero energy. Each band is spin degenerate which enhances the $S_z$ symmetry to a full $SU(2)$ and consequently, the state is paramagnetic. The right hand side of the figure shows the Fermi surface (blue, dashed) where we have kept the electron density fixed at the half-filling. Consistent with the Oshikawa's argument \cite{Oshikawa00a}, one finds that the Fermi surface is `large', and occupies half of the Brillouin zone which is depicted by the shaded area in Fig.\ref{fig:fullbandstructure}. The effective chemical potential required for the half filling is marked by the dashed orange line in the middle panel.

Finally, one may also consider hybridization of the form $\sum_{\< \pmb{x}, \pmb{i}\>} \cdag{\pmb{x},\uparrow}\fd{\pmb{i},\uparrow} -(-1)^{\pmb{x}}\cdag{\pmb{x},\downarrow}\fd{\pmb{i},\downarrow} +h.c.\,$ which preserves the particle-hole symmetry listed as ($\sigma^x$, -) in Table \ref{Tab:PHsym}, but still breaks $(\sigma^z,-), (\sigma^0,+)$ and $(\sigma^0,-)$, as well as $TR_1$ and $TR_2$. This mean-field is also motivated by the structure of our Hamiltonian where the Kondo interaction has an additional sign that depends on the sublattice. Due to time-reversal symmetry breaking, such a state will generically result in magnetization along the z-direction. 

\section*{II.~~~Details on the Method}

Let us first write down the fermionized  Hamiltonian that is simulated, $\hat{H}_{\mathrm{qmc}}$, and then  show its equivalence to  Eq.~(\ref{KondoH}).

\begin{widetext}
\begin{eqnarray}
	\hat{H}_{\mathrm{qmc}} &=&-t \sum_{\< \pmb{x}, \pmb{y}\>,\sigma}\ctdag{\pmb{x},\sigma}\ctd{\pmb{y},\sigma} + h.c. 
	-\frac{J^{\perp}}{4} \sum_{\< \pmb{i}, \pmb{j}\>}
	\left[
		2\left(
		\sum_{\sigma}\ftdag{\pmb{i},\sigma}\ftd{\pmb{i},\sigma} + h.c. 
		\right)^2
		+\left( \vphantom{\sum_{\sigma}}
		n^{\tilde{f}}_{ \pmb{i}} + n^{\tilde{f}}_{ \pmb{j}} - 1
		\right)^2
	\right]
	\nonumber \\ && 
	-\frac{J^{z}}{4}  \sum_{\hexagon} \sum_{\pmb{i}_{\hexagon} < \pmb{j}_{\hexagon}}\left(
	n^{\tilde{f}}_{ \pmb{i}} - n^{\tilde{f}}_{ \pmb{j}}
	\right)^2
	 -\frac{J_K}{4} \sum_{\< \pmb{i}, \pmb{x}\>}\left(
	\sum_{\sigma}\ftdag{\pmb{i},\sigma}\ctd{\pmb{x},\sigma} + h.c. 
	\right)^2
	 \, , \label{eq:Hqmc}
\end{eqnarray}
\end{widetext}
with $(\ctdag{\pmb{x},\uparrow},\ctdag{\pmb{x},\downarrow})=(\cdag{\pmb{x},\uparrow},(-1)^{\pmb{x}}\cd{\pmb{x},\downarrow})$ and $(\ftdag{\pmb{i},\uparrow},\ftdag{\pmb{i},\downarrow})=(\fdag{\pmb{i},\uparrow},\fd{\pmb{i},\downarrow})$. The Hamiltonian above is identical to Eq.~(\ref{KondoH}) up to the following five terms in $\hat{H}_{\mathrm{qmc}} - \hat{H}$.
The first term 
$+ (J^\perp+4J^z)\sum_{\pmb{i}} ( n^f_{\pmb{i}}-1 ) ^ 2$
is the well known repulsive Hubbard interaction that suppress charge fluctuations. The local parity of the $f$-electrons $( n^f_{\pmb{i}}-1 ) ^ 2$ commutes with the Hamiltonian as the relevant terms 
$+ J^\perp \sum_{\left\langle \pmb{i},\pmb{j} \right\rangle} \fdag{\pmb{i},\uparrow}  \fdag{\pmb{i},\downarrow}  \fd{\pmb{j},\downarrow}  \fd{\pmb{j},\uparrow}  + h.c.$ 
and 
$+ \frac{J_K}{2} \sum_{\left\langle \pmb{i},\pmb{x} \right\rangle} (-1)^{\pmb{x}}  \fdag{\pmb{i},\uparrow}  \fdag{\pmb{i},\downarrow}  \cd{\pmb{x},\downarrow}  \cd{\pmb{x},\uparrow}  + h.c.$
modify the local occupation by 2. Hence the Hubbard interaction  projects onto the  sector with singly occupied $f$-electron sites exponentially fast and  the relevant scale is set by $\beta(J^\perp+4J^z)$. In this subspace, all other contributions of
$+ \frac{J^\perp}{2}  \sum_{\left\langle \pmb{i},\pmb{j} \right\rangle}  (n^f_{\pmb{i}}-1)(n^f_{\pmb{j}}-1)$ and $+ \frac{J_K}{4}  \sum_{\left\langle \pmb{i},\pmb{x} \right\rangle}  (n^f_{\pmb{i}}-1)(n^c_{\pmb{x}}-1)$ 
, vanish such that $\hat{H}_{\mathrm{qmc}}|_{( n^f_{\pmb{i}}-1 ) ^ 2=0} = \hat{H}$. The interested reader is referred to the supplemental material, i.e. Eq.~(9), of Ref.~\cite{SatoT17_1}.

The efficient projection due to the repulsive Hubbard interaction however also introduces a challenge for the numerical stability of the algorithm. Here we have to control the various scales of $A_j=\prod_{i=0}^{j}B_i$ where $B_i$ is the product of all exponentiated operators on the $i$th time slice. Apparently, this model generated Eigenvalues in $A_j$ which exceeded the range of double precision which is of order $10^{\pm 308}$. To overcome this issue, we implemented the following stabilization scheme. Assume that we already have a $QR$ decomposition of $A_{j-1}=Q_{j-1}e^{\lambda_{j-1}}R_{j-1}$ where $Q_{j-1}$ is the orthogonal part, $e^{\lambda_{j-1}}$ is diagonal and separates the main scales, and $R_{j-1}$ contains the mixing of them. To generate $A_j=B_j A_{j-1}$ we perform the following steps:

\begin{enumerate}
	\item Calculate $M_j= B_j Q_{j-1}$
	\item Use the permutation $P_j$ to sort the columns of $M_j=\tilde{M}_j P_j$ according to the column norm of $M_j e^{\lambda_{j-1}}$. Permute $\lambda_{j-1}$ and $R_{j-1}$ with  $P^{-1}_j$ to correct this manipulation
	\item Perform a $QR$ decomposition of $M_j=Q_j\tilde{R}_j$ without pivoting.
	\item Extract the scales of $\tilde{R}$ as $(D_j)_n = |(\tilde{R}_j)_{nn}|$.
	\item Determine the new scales $\lambda_{j} = \log(D_j) + \lambda_{j-1}$.
	\item Calculate $R_j=D^{-1}_j e^{-\lambda_{j-1}} \tilde{R}_j e^{\lambda_{j-1}} R_{j-1}$
\end{enumerate}

This scheme keeps all the advantages of $QR$ decomposition with pivoting to handle exponentially large and small scales of $A_j$ which is paramount to a stable BSS algorithm, even when double precision suffices. Here, we did not store the scales as $D$'s but rather as $e^{\lambda_{j-1}}$ to handle numbers much larger than $10^{\pm 308}$.

\section*{III.~~~Time displaced Greens function}

\begin{figure}
	\centering
	\includegraphics[width=0.4925\linewidth]{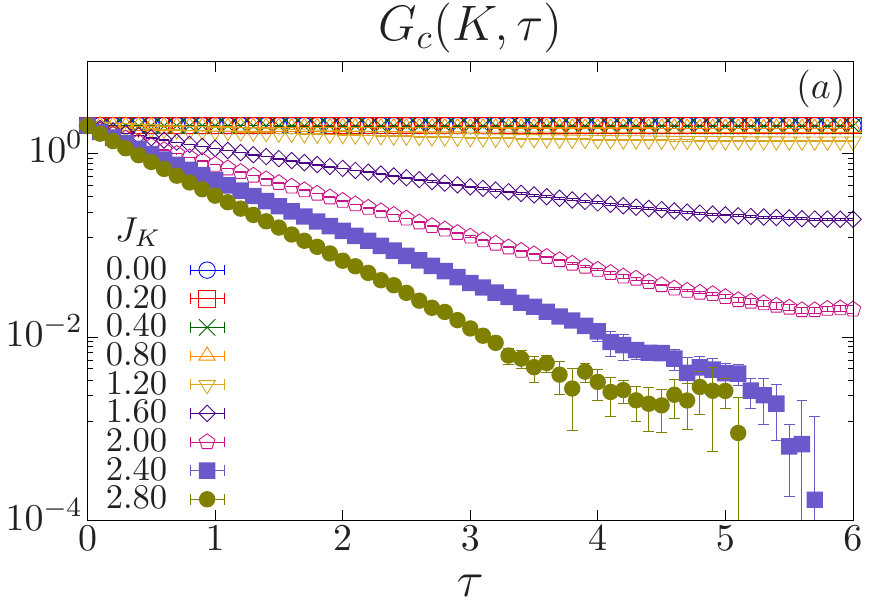} \hfill
	\includegraphics[width=0.4925\linewidth]{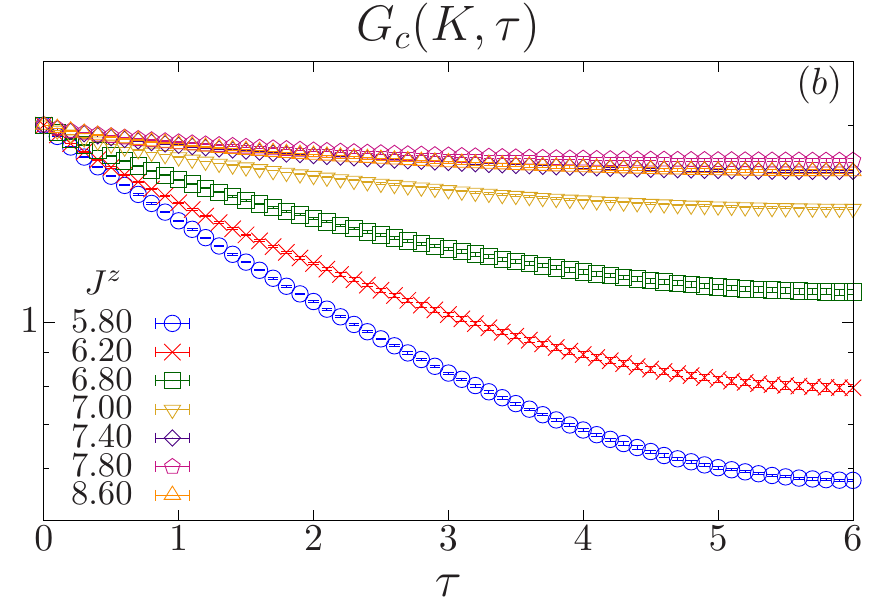}\\
	\includegraphics[width=0.4925\linewidth]{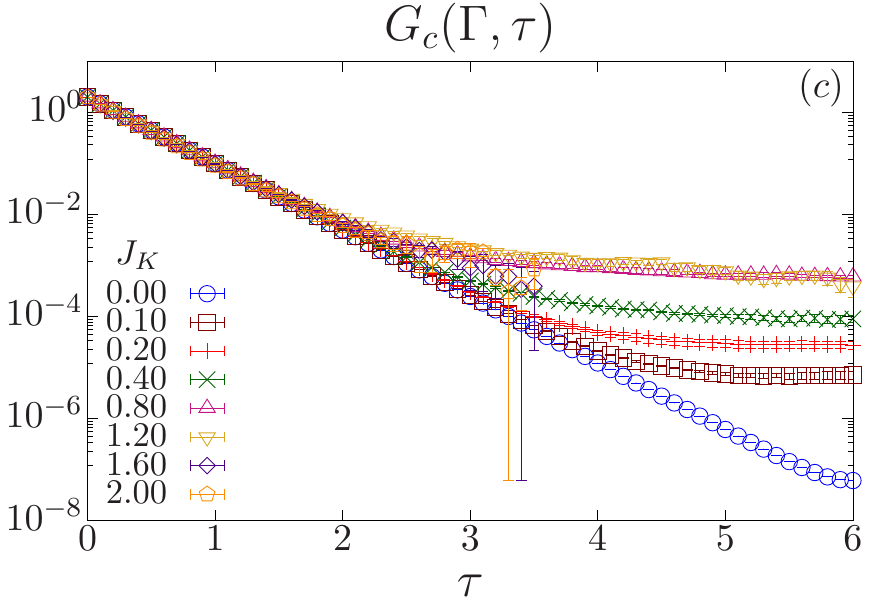} \hfill
	\includegraphics[width=0.4925\linewidth]{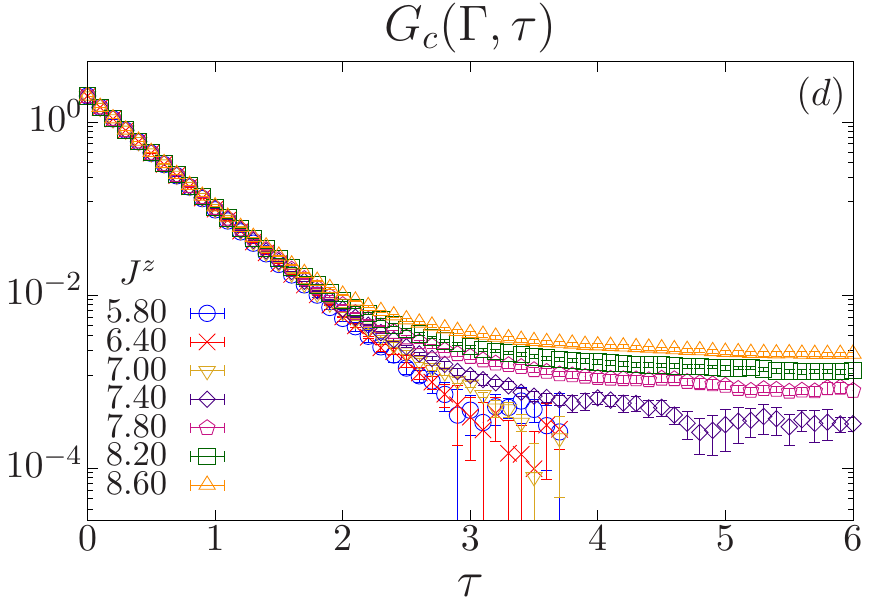}\\
	\includegraphics[width=0.4925\linewidth]{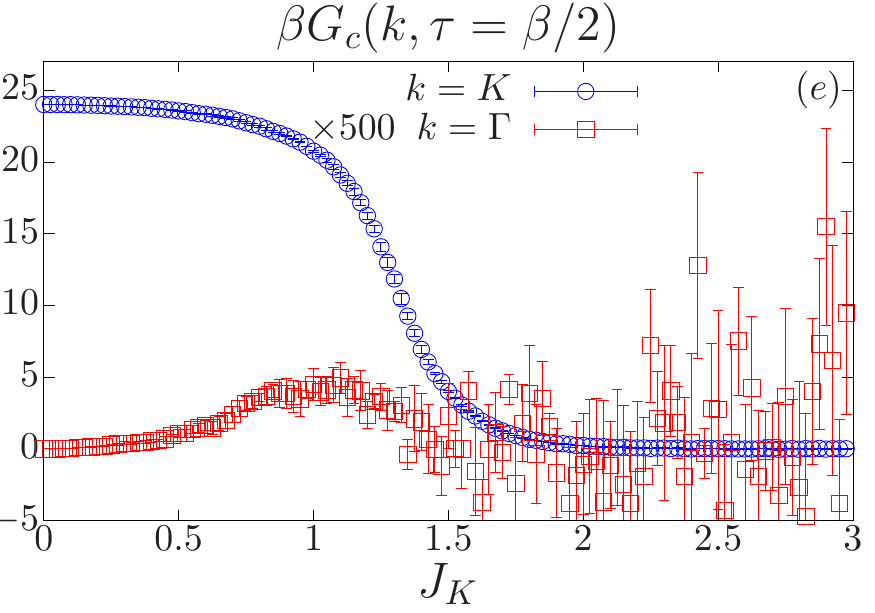} \hfill
	\includegraphics[width=0.4925\linewidth]{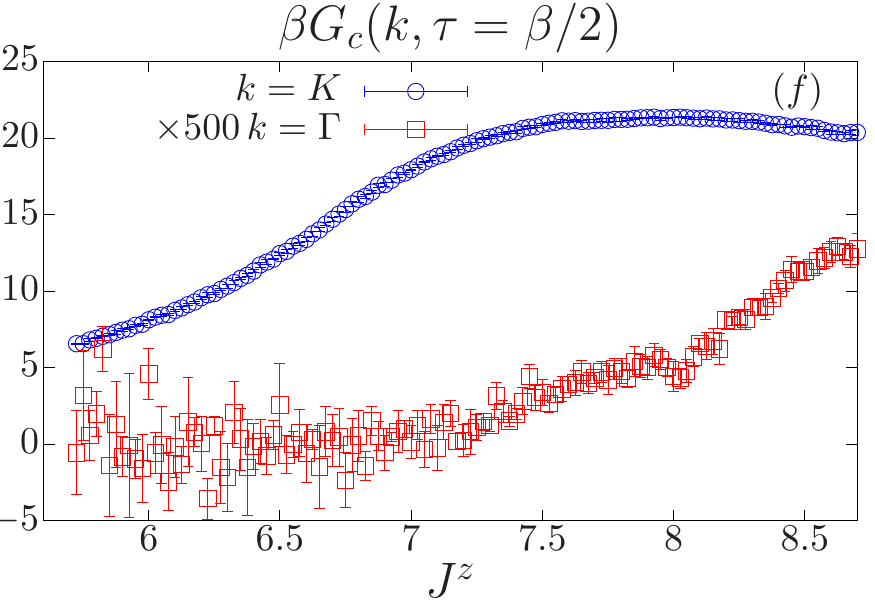}
	\caption{ (color online)
		The simulations were performed on  the $L=3\times 6$ lattice at an inverse temperature of $\beta =12$. Left panels corresponds to the $J_K$ scan at  $J^z=7.5$  and  the right to the $J^z$ scan at  $J_K=1.0$. For large $J_K$ or small $J^z$, we restricted the time domain in (c) and (d) to $\tau<3.5$ and $\tau<3.75$, respectively, since beyond this scale, the data becomes very noisy.
		\label{fig:beta_half}
	}
\end{figure}

Here we provide the imaginary  time displaced Greens functions of conduction electrons,
$G_c(\pmb{k},\tau) =   \sum_{\alpha, \sigma}  \langle \hat{c}^{\dagger}_{{\pmb k}, \alpha, \sigma} (\tau)  \hat{c}^{\phantom\dagger}_{{\pmb k}, \alpha,\sigma} (0)  \rangle  $ where $\alpha$ is the orbital and $\sigma$ the spin  index.      The dynamical data presented in the main text, is obtained  by solving
\begin{equation}
	  G_c(\pmb{k},\tau)     = \frac{1}{\pi} \int d \omega   \frac{ e^{- \tau \omega} }{ 1 + e^{-\beta \omega}}   A_c(\pmb{k}, \omega)
\end{equation}
for  $A_c(\pmb{k}, \omega)$ using the stochastic maximum entropy method \cite{Sandvik98,Beach04a}.   The features present in the dynamical data can  clearly be detected in the imaginary time data  which we report in this section. 
In \fig{fig:beta_half}, the left hand side panels presents the  $J_K$  scan at a fixed $J^z=7.5$ whereas on the right hand side we  show the $J^z$ scan  at  constant $J_K=1.0$. 
 
Panels (a) and (b) depicts the Greens function at the Dirac points.  In both cases, the gapless mode is clearly visible  in the  FL* phase  since  $G_c(\pmb{K},\tau) $    shows a plateau at 
{\it large}  imaginary times.  This height of the plateau corresponds to the quasi-particle residue.   
  
Panels (c)  and (d) present the equivalent data   but at the $\Gamma$ point.  In the FL* phase we see a  clear  feature  with small intensity at {\it large}  values of $\tau$. It is this feature in the imaginary time Green function that generates the low energy spectral weight  in  \fig{fig:Z2_Jscan}(d) and \fig{fig:Z2_Vscan}(d) in the FL* phase.    As mentioned in the article, we interpret this feature as a signature of the vison excitation. 
 
Another possible analysis stems from the identity,
\begin{equation}
 \lim_{\beta \rightarrow  \infty } \beta G_c(\pmb{k}, \tau=\beta/2)  = A_c(\pmb{k}, \omega =0),    
\end{equation}
that holds provided that $A_c(\pmb{k}, \omega) $ is a smooth function.     At finite values of  $\beta$,   $\beta G_c(\pmb{k}, \tau=\beta/2)$   will  provide an estimate of the spectral weight in  an energy window around $\omega =0 $ of width set by $1/\beta$.    Panels (e) and (f) plot this quantity both at the $\Gamma$  and Dirac points.  Overall, these panels again confirm  
that in the FL* phase we observe  low energy excitations  with small intensity at the $\Gamma$ point   and low energy excitations with large spectral weight at the Dirac point.  Note that in panel (e), corresponding to the $J_K$ scan, the intensity of the feature at the $\Gamma$ point  first grows and then decreases since both at $J_K=0$, where the spin and conduction  electrons decouple and the conduction electrons form a Dirac spectrum,  and at $J_K \gg 1$ where in the magnetic insulating phase, no  low lying single particle weight is expected at the $\Gamma$ point. 

\end{document}